\documentclass[12pt,a4paper,twoside]{article}
\newenvironment{changemargin}[2]{%
\begin{list}{}{%
\setlength{\leftmargin}{#1}%
\setlength{\rightmargin}{#2}%
}%
\item[]}
{\end{list}}
\usepackage{graphicx}
\usepackage{color}
\usepackage{lscape}
\usepackage{hyperref}
\usepackage{epstopdf}
\usepackage{lscape}
\usepackage{amsmath}
\usepackage{hyperref}
\usepackage{amsmath}
\usepackage{setspace}
\begin{document}
\baselineskip=0.30in
{\bf \LARGE

\begin{changemargin}{-1.2cm}{0.5cm}
\begin{center}
{Spectroscopic study of some diatomic molecules via the proper quantization rule}
\end{center}

\end{changemargin}
}
\vspace{4mm}
\begin{center}
{\Large{\bf Babatunde J. Falaye $^a$$^{,}$$^\dag$$^{,}$}}\footnote{\scriptsize E-mail:~ fbjames11@physicist.net;~ babatunde.falaye@fulafia.edu.ng\\ \dag{Corresponding} author}\Large{\bf ,} {\Large{\bf Sameer M. Ikhdair $^b$$^{,}$}}\footnote{\scriptsize E-mail:~ sameer.ikhdair@najah.edu;~ sikhdair@gmail.com.} \Large{\bf and} {\Large{\bf Majid Hamzavi $^c$$^{,}$}}\footnote{\scriptsize E-mail:~ majid.hamzavi@gmail.com }
\end{center}
{\small
\begin{center}
{\it $^\textbf{a}$Applied Theoretical Physics Division, Department of Physics, Federal University Lafia,  P. M. B. 146, Lafia, Nigeria.}
{\it $^\textbf{b}$Department of Physics, Faculty of Science, an-Najah National University, New campus, P. O. Box 7, Nablus, West Bank, Palestine.}
{\it $^\textbf{c}$Department of Physics, University of Zanjan, Zanjan, Iran.}
\end{center}}

\begin{center}
\textbf{J. Math. Chem. (2015) DOI 10.1007/s10910-015-0491-9}
\end{center}

\begin{abstract}
\noindent
Spectroscopic techniques are very essential tools in studying electronic structures, spectroscopic constants and energetic properties of diatomic molecules. These techniques are also required for parametrization of new method based on theoretical analysis and computational calculations. In this research, we apply the proper quantization rule in spectroscopic study of some diatomic molecules by solving the Schr\"{o}dinger equation with two solvable quantum molecular systems-Tietz-Wei and shifted Deng-Fan potential models for their approximate nonrelativistic energy states via an appropriate approximation to the centrifugal term. We show that the energy levels can be determined from its ground state energy. The beauty and simplicity of the method applied in this study is that, it can be applied to any exactly as well as approximately solvable models. The validity and accuracy of the method is tested with previous techniques via numerical computation for H$_2$ and CO diatomic molecules. The result also include energy spectrum of 5 different electronic states of NO  and 2 different electronic state of ICl.
\noindent
\end{abstract}

{\bf Keywords}: Proper quantization rule; formula method; Schr\"{o}dinger equation; Tietz-Wei diatomic molecular potential; Shifted Deng-Fan diatomic molecular potential.

{\bf PACs No.}: 003.65.Fd, 03.65.Ge, 03.65.Ca, 03.65-W

\section{Introduction}
\label{sec1}
The exact solutions of solvable quantum potential models have received much interest since they provide us some insight into the physical problem under consideration. Over the past years, various eigensolution techniques have been proposed to solve quantum potential models. Few of these methods are: formula method \cite{NEQ1}, Nikiforov-Uvarov method \cite{SM1}, the asymptotic iteration method \cite{SM2}, the supersymmetric quantum mechanics \cite{SM3}, the factorization method \cite{SM4}, wave function ansatz method \cite{SM5}, the generalized pseudospectral method (GPS) \cite{P1} and the exact quantization rule (EQR) \cite{SM7,B1,B2}. Notes on these techniques can be found in ref. \cite{NEQ2}

Recently, the EQR has been proposed to solve the wave equations with some exactly or approximately solvable quantum potentials for their energy eigenvalues and wave functions. Nevertheless, such solutions involve highly complicated integral calculations, in particular when calculating the quantum correction term. Therefore, in order to avoid these difficulties, Serrano et al have proposed a new way to treat these problems and called it the proper quantization rule (PQR)  \cite{SM10}. Furthermore, it has been shown that PQR is a powerful tool in finding the eigenvalues for all solvable quantum systems \cite{SM11,SM13}.

Furthermore, the study of the bound state processes is fundamental to understanding the molecular spectrum of the diatomic molecules and their properties in quantum mechanics. In light of this, there has been a growing interest in searching for the empirical potential functions for diatomic molecules in chemical physics and related areas \cite{BE1,SM14}.  The reason is that such potentials provide the compact way to summarize what we know about a molecule. Thus, efforts to construct a universal potential function that fit experimental data in computational chemistry have been made by many researchers. It has been found that the exponential type molecular potentials are better than the harmonic oscillator in simulating the atomic interaction for diatomic molecules. 

In this context, to achieve the goal of the present work, we study the spectrum of some diatomic molecules using two exponential-type of molecular models; namely, the Tietz-Wei and shifted Deng-Fan potential models \cite{SM14, SM15}. The bound state solution of the Schr\"odinger equation with these diatomic molecular potentials provides the rotational-vibrational energy states of the diatomic molecules in an accurate manner. We apply PQR to obtain the energy spectrum of the two molecular potential models and then obtain the rotational-vibrational energy states for various diatomic molecules. 

This work is organized as follows. In section 2, we briefly introduce the PQR. In section 3, we apply the method to obtain the energy spectrum of the Schr\"{o}dinger equation with Tietz-Wei and shifted Deng-Fan molecular potentials. We give our numerical results and discussions in section 4. Some concluding remarks are given in section 5.  
        
\section{A brief review to proper quantization rule}
In this section, we give a brief review to this method \cite{SM7, SM10,SM11}. The one dimensional Schr\"{o}dinger equation takes the form:
\begin{equation} 
\frac{d^2}{dx^2}\psi(x)+\frac{2\mu}{\hbar^2}\left[E-V(x)\right]\psi(x)=0,
\label{E1}
\end{equation}
and can be re-written as
\begin{equation}
\phi'(x)+\phi(x)^2+k(x)^2=0,\ \ \ \mbox{with}\ \ \ \ \ \ k(x)=\sqrt{\frac{2\mu}{\hbar^2}[E-V(x)]},
\label{E2}
\end{equation}
where $\phi(x)=\psi'(x)/\psi(x)$ is the logarithmic derivative of the wave function $\psi(x)$. The prime denotes the derivative with respect to the variable $x$. $\mu$ denotes the reduced mass of the two interacting particles. $k(x)$ is the momentum and $V(x)$ is a piecewise continuous real potential function of $x$. According to Yang \cite{BJ5} ``For the Sturm-Liouville problem, the fundamental trick is the definition of a phase angle which is monotonic with respect to the energy''\cite{BJ5}. Thus, for the Schr\"{o}dinger equation, the phase angle is the logarithmic derivative $\phi(x)$. From equation (\ref{E2}), as $x$ increases across a node of wave function $\psi(x)$, $\phi(x)$ decreases to $-\infty$, jumps to $+\infty$ and then decreases again.  

In 2005, Ma and Xu \cite{SM7} by carefully studying one-dimensional Schr\"{o}dinger equation generalized this exact quantization rule to the 3D radial Schr\"{o}dinger equation
with spherically symmetric potential by simply making the replacements $x\rightarrow r$ and $V(x)\rightarrow V_{eff}(r)$:
\begin{equation}
\int_{r_a}^{r_b}k(r)dr=N\pi+\int_{r_a}^{r_b}\phi(r)\left[\frac{dk(r)}{dr}\right]\left[\frac{d\phi(r)}{dr}\right]^{-1}dr,\ \ \ \ k(r)=\sqrt{\frac{2\mu}{\hbar^2}[E-V_{eff}(r)]}, 
\label{E3}
\end{equation}
where $r_A$ and $r_B$ are two turning points determined by $E=V_{eff}(r)$. The $N=n+1$ is the number of the nodes of $\phi(r)$ in the region $E_{n\ell}=V_{eff}(r)$ and is larger by one than the number $n$ of the nodes of wave function $\psi(r)$.  The first term $N\pi$ is the contribution from the nodes of the logarithmic derivative of wave function, and the
second is called the quantum correction.  Ma and Xu \cite{SM7} found that for all well-known exactly solvable quantum systems, this quantum correction is independent of the number of nodes of wave function. Accordingly, it is enough to consider the ground state in calculating the quantum correction $Q_c=\int_{r_A}^{r_B}k_0'(r)\frac{\phi_0}{\phi_0'}dr$.

The integrals in equations (\ref{E3}) and the calculation of quantum correction term are not easy to obtain for various quantum mechanical problems. This motivated Serrano et al in 2010 to propose Qiang-Dong proper quantization rule \cite{SM10}, so as to simplify the quantum correction terms. This rule can be summarized as follows:
\begin{equation}
\int_{r_{a}}^{r_{b}}k(r)dr=\int_{r_{0a}}^{r_{0b}}k_0(r)dr+n\pi \ \ \ \ \mbox{with}\ \ \ \ n=N-1.
\label{E4}
\end{equation}
In the approach, it is required to first calculate the integral on the LHS of equation (\ref{E4}) and then replace energy levels $E_n$ in the result by the ground state energy $E_0$ to obtain the second integral (RHS). This quantization rule has been used in many physical systems to obtain the  exact solutions of many exactly solvable quantum systems \cite{SM7,SM10,SM11,BJ6,BJ7}
\section{Application to some diatomic molecular potentials}
In this section, we apply the Qiang-Dong proper quantization to study the rotation vibrational of some diatomic molecular potentials. Also, where necessary, we compare our results with the ones obtained before in the literature.
\subsection{Tietz-Wei molecular potential}
The Tietz-Wei diatomic molecular potential we examine in this section is defined as \cite{NEQ2,BJ8,BJ9}
\begin{equation}
V(r)=D\left[\frac{1-e^{-b_h(r-r_e)}}{1-c_he^{-b_h(r-r_e)}}\right]^2,
\label{E5}
\end{equation}
with $b_h=\delta(1-c_h)$, $r_e$ is the molecular bond length, $\delta$ is the Morse constant (denoted as $\beta$ in some other research papers), $D$ is the potential well depth and $c_h$ is an optimization parameter obtained from ab initio or Rydberg-Klein-Rees (RKR) intramolecular potentials. $r$ is the internuclear distance. When the potential constant approaches zero, i.e. $c_h\rightarrow 0$, the TW potential reduces to the Morse potential \cite{BJ10}. The shape of this potential is shown in Figure \ref{fig1} for different molecules. To study any quantum physical model characterized by the diatomic molecular potential given by equation (\ref{E4}), we need to solve the following Schr\"{o}dinger equation:
\begin{figure}[!htb]
\centering\includegraphics[height=85mm,width=150mm]{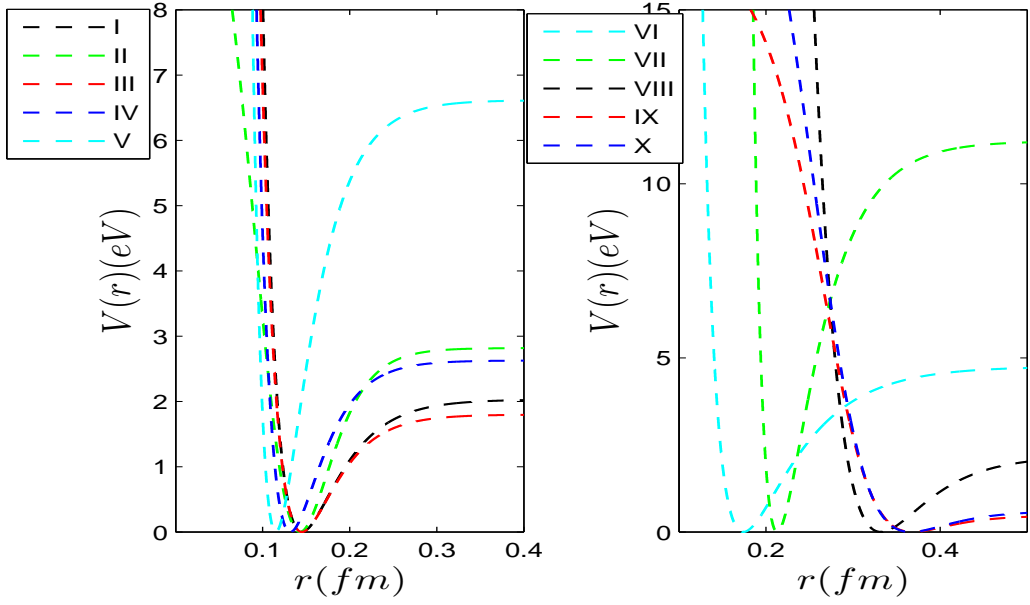}
\caption{{\protect\footnotesize Shape of Tietz-Wei diatomic molecular potential for different diatomic molecules: (I) NO$\left(a^4\Pi_i\right)$ (II) NO$\left(B^2\Pi_r\right)$ (III) NO$\left(L'^2\phi\right)$ (IV) NO$\left(b^4\Sigma^{-}\right)$ (V) NO$(X^2\Pi_r)$ (VI) H$_2\left(X^1\Sigma^+_g\right)$ (VII) CO$\left(X^1\Sigma^+\right)$ (VIII) ICl$\left(X^1\Sigma_g^{+}\right)$ (IX) ICl$\left(A^3\Pi_1\right)$ (X) ICl$\left(A'^3\Pi_2\right)$.}}
\label{fig1}
\end{figure}
\begin{equation}
\left(\frac{P^2}{2m}+V(r)-E_{n\ell}\right)\psi_{n,\ell,m}(r,\theta,\phi)=0.
\label{E6}
\end{equation}
In this section, we take the $V(r)$ as the Tietz-Wei potential. Now we begin by applying the method of variable separation so as to split equation (\ref{E6}) into radial and angular part. Thus, by taking the wavefunction $\psi_{n,\ell,m}(r,\theta,\phi)$ as $r^{-1}R_{n\ell}(r)Y_{\ell m}(\theta,\phi)$ the radial part can be found as
\begin{equation}
\frac{d^2R_{n\ell}(r)}{dr^2}+\frac{2\mu}{\hbar^2}\left[E_{n\ell}-D\left[\frac{1-e^{-b_h(r-r_e)}}{1-c_he^{-b_h(r-r_e)}}\right]^2-\frac{\ell(\ell+1)\hbar^2}{2\mu r^2}\right]R_{n\ell}(r)=0,
\label{E7}
\end{equation}
where $n$, $\ell$ and $E_{n\ell}$ denote the principal quantum numbers, orbital angular momentum numbers  and the bound state energy eigenvalues of the system under consideration (i.e., $E_{n\ell}<0$ ),  respectively. It is generally known that for $\ell=0$, problem (\ref{E7}) is exactly solvable but for $\ell\neq0$, it isn't. Therefore, in order to solve the above equation for $\ell\neq0$ states, Hamzavi et al \cite{BJ9} found that the following formula
\begin{figure}[!t]
 \includegraphics[height=85mm,width=150mm]{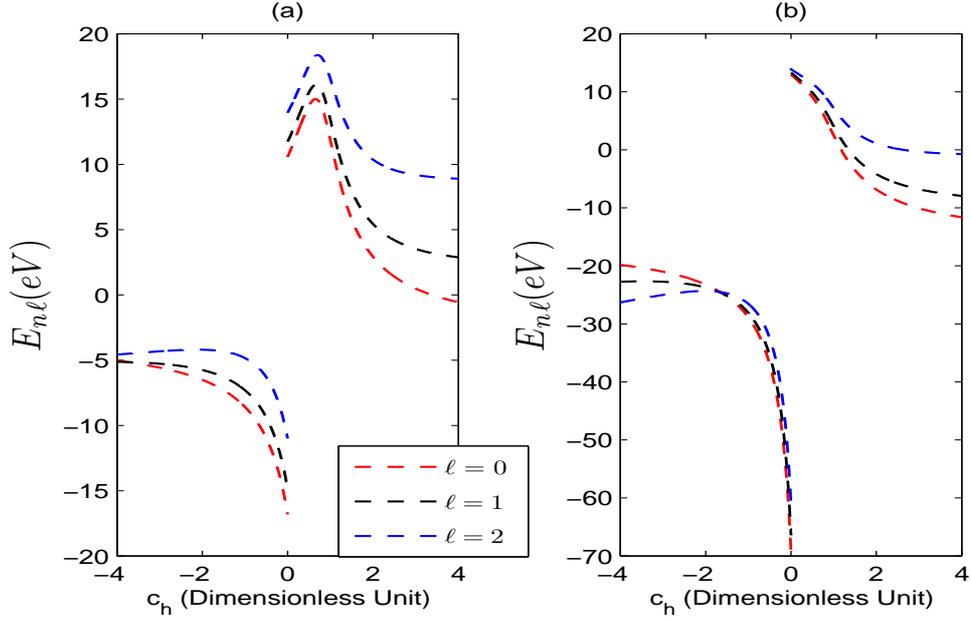}
\caption{{\protect\small (a) The variation of the ground state energy spectrum for various values of $\ell$ as a function of the potential constant $c_h$. We choose $\mu=1$, $b_h=5$,  $r_e=0.8$ and $D=15$. (b) The variation of the first excited energy state for various $\ell$ as a function of the potential constant $c_h$.}}
\label{fig2}
\end{figure}
\begin{figure}[!t]
 \includegraphics[height=85mm,width=150mm]{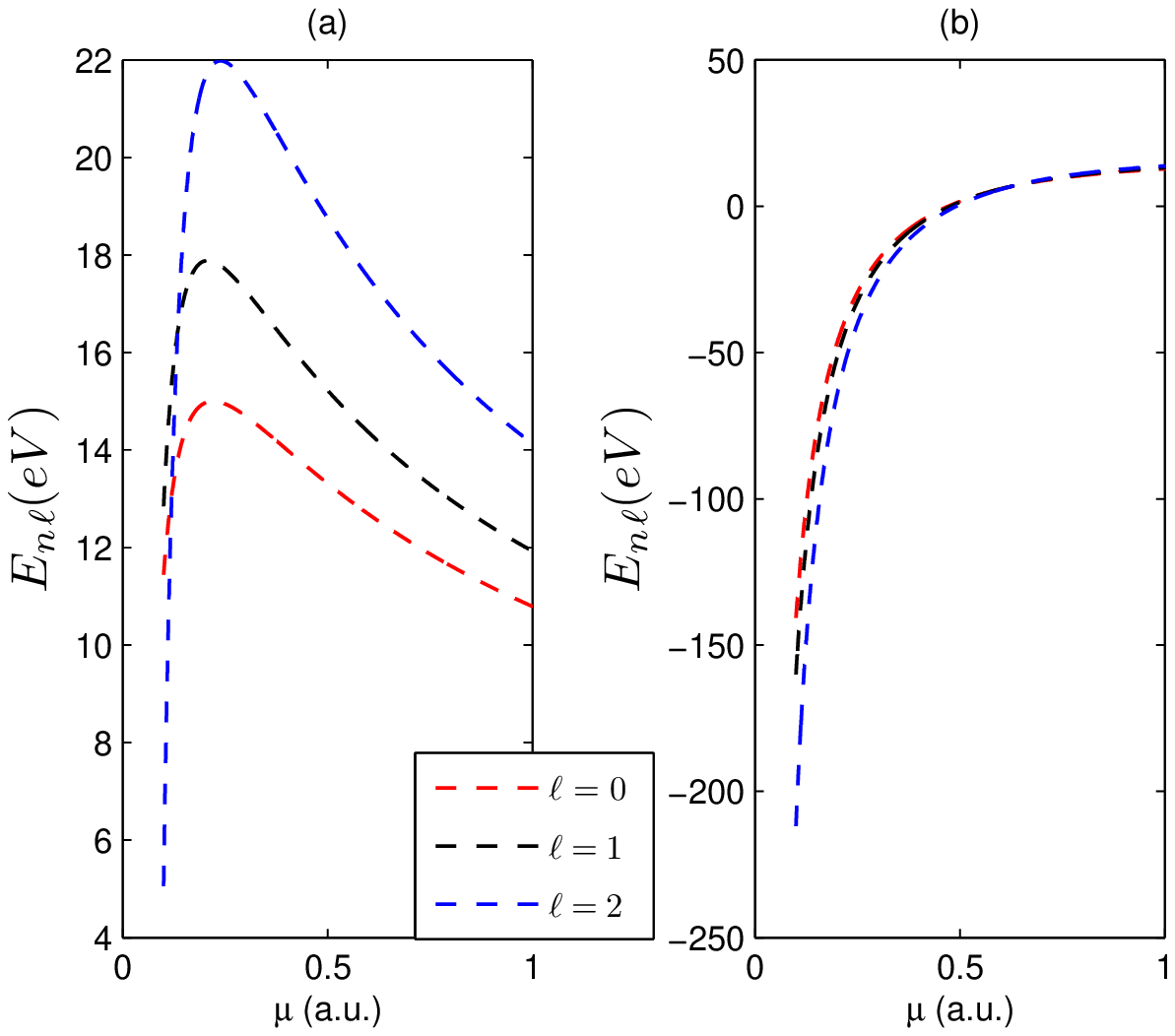}
\caption{{\protect\small (a) The variation of the ground state energy state for various values of $\ell$ as a function of the particle mass $\mu$. We choose $c_h=0.03$, $b_h=5$,  $r_e=0.8$ and $D=15$. (b) The variation of the first excited energy state for various $\ell$ as a function of the particle mass $\mu$.}}
\label{fig3}
\end{figure}
\begin{figure}[!t]
\centering \includegraphics[height=85mm,width=150mm]{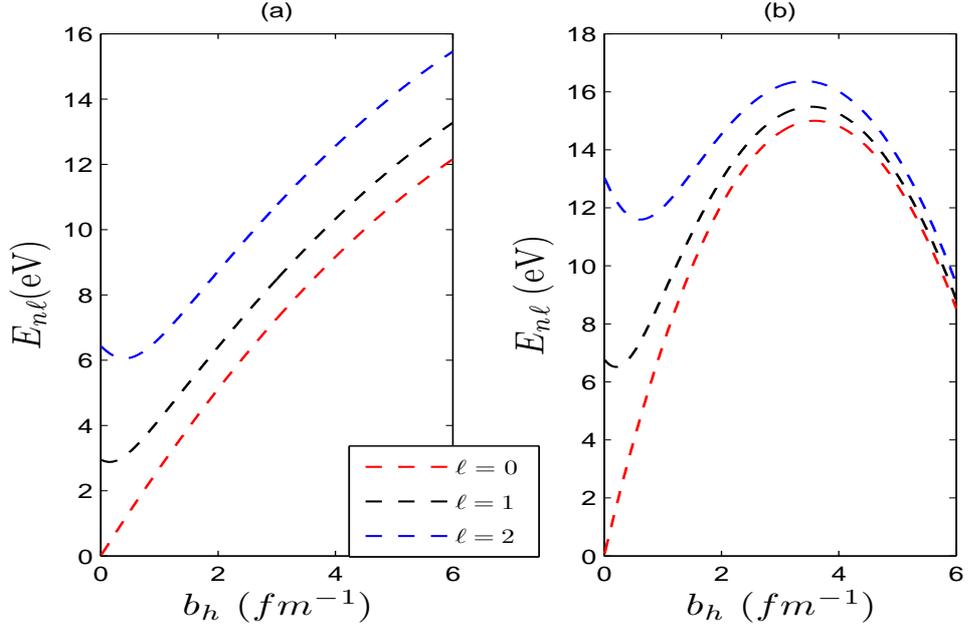}
\caption{{\protect\small (a) The variation of the ground state energy state for various values of $\ell$ as a function of the parameter $b_h$. We choose $c_h=0.03$, $\mu=1$, $r_e=0.8$ and $D=15$. (b) The variation of the first excited energy state for various $\ell$ as a function of the parameter $b_h$.}}
\label{fig4}
\end{figure}
\begin{figure}[!t]
\includegraphics[height=85mm,width=150mm]{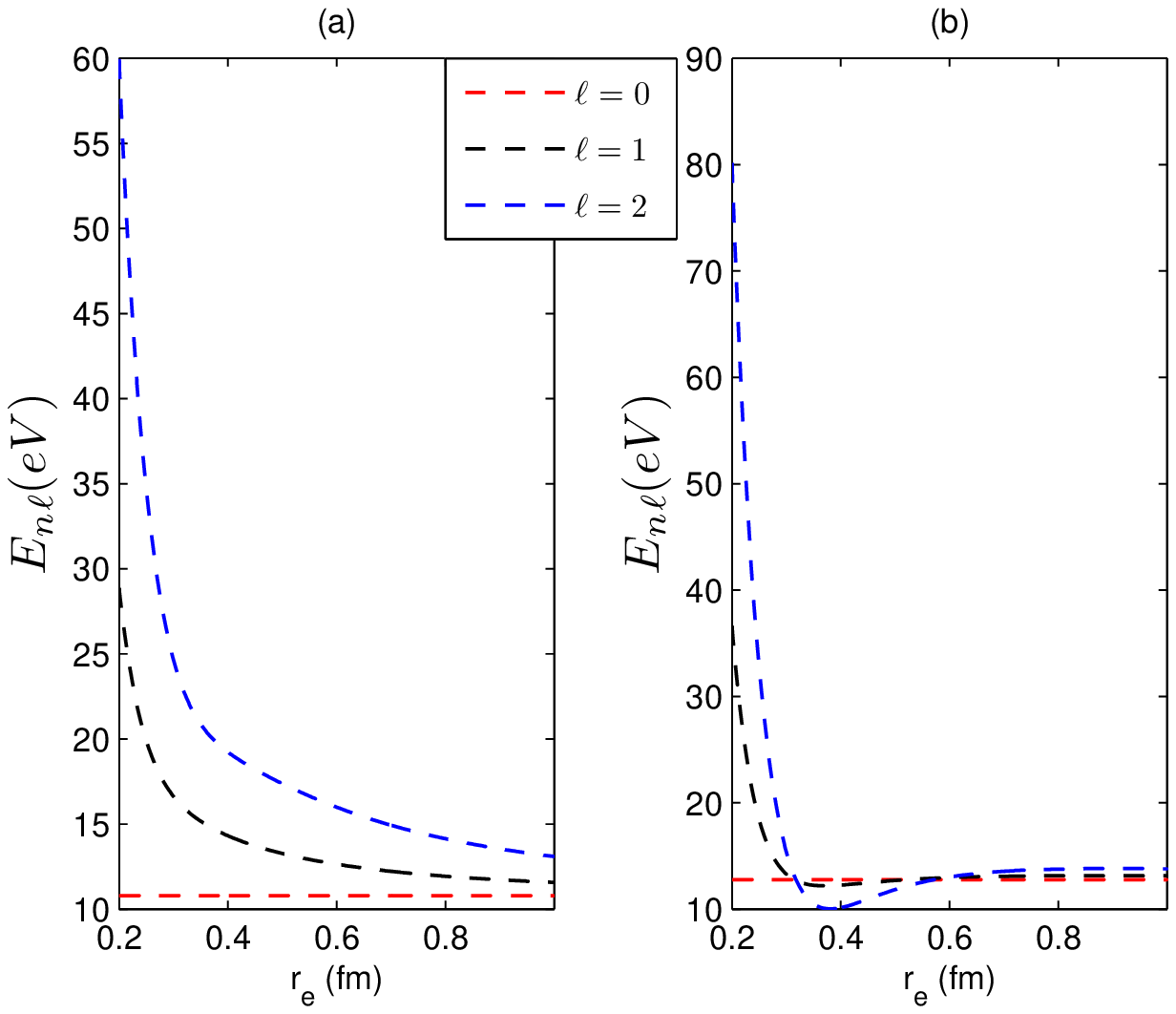}
\caption{{\protect\small (a) The variation of the ground state energy state for various values of $\ell$ as a function of the molecular bond length $r_e$. We choose $\mu=1$, $b_h=5$,  $c_h=0.03$ and $D=15$. (b) The variation of the first excited energy state for various $\ell$ as a function of the molecular bond length $r_e$}}
\label{fig5}
\end{figure}
\begin{figure}[!t]
\includegraphics[height=85mm,width=150mm]{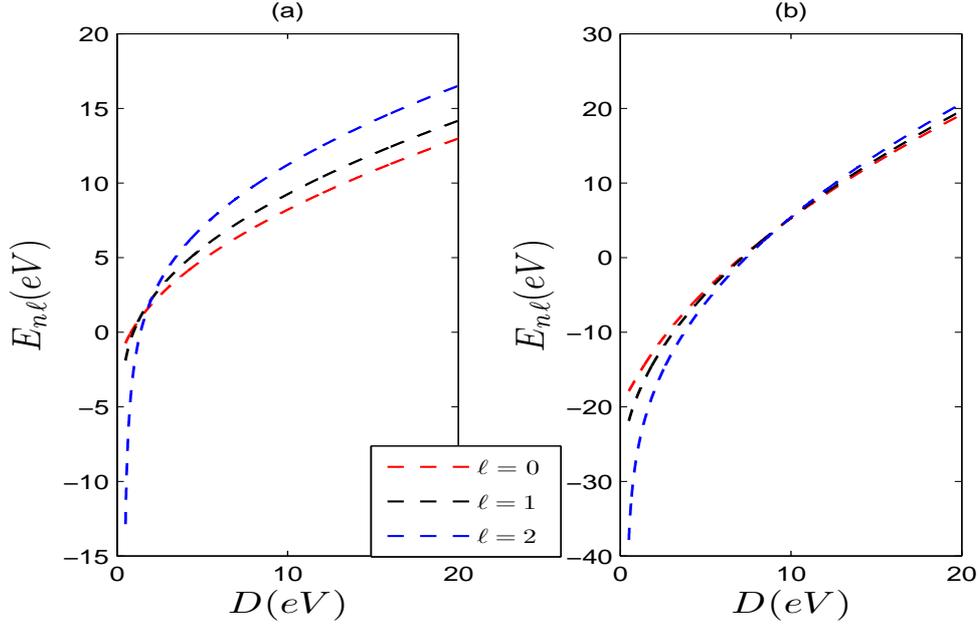}
\caption{{\protect\small (a) The variation of the ground state energy state for various values of $\ell$ as a function of the potential well depth $D$. We choose $\mu=1$, $b_h=5$,  $c_h=0.03$ and $r_e=0.8$. (b) The variation of the first excited energy state for various $\ell$ as a function of the potential well depth $D$}}
\label{fig6}
\end{figure}
\begin{table}[!t]
{\scriptsize
\caption{Model parameters of the diatomic molecules studied in the present work. }\vspace*{10pt}
\begin{tabular}{cccccccc}\hline\hline
{}&{}&{}&{}&{}&{}&{}&{}\\[-1.0ex]
Molecules(states)&$c_h$&$\mu/ 10^{-23}(g)$& $b_h(\AA^{-1})$&$r_e (\AA)$&$D(cm^{-1})$&$\beta (\AA^{-1})$&Refs\\[2.5ex]\hline\hline
NO$\left(a^4\Pi_i\right)$	&0.0082003	&1.249	&2.408413	&1.451	&16361&2.428326&\cite{BJ14}	\\[1ex]
NO$\left(B^2\Pi_r\right)$	&-0.482743	&1.249	&3.42650	&1.428	&22722&2.310923&\cite{BJ14}\\[1ex]
NO$\left(L'^2\phi\right)$	&-0.073021	&1.249	&2.73796	&1.451	&14501&2.551645&\cite{BJ14}\\[1ex]
NO$\left(b^4\Sigma^{-}\right)$	&-0.085078	&1.249	&3.01538	&1.318	&21183&2.778957&\cite{BJ14}	\\[1ex]
NO$(X^2\Pi_r)$&0.013727&1.249	&2.71559	&1.151	&53341&2.7534&\cite{BJ14}\\[1ex]
H$_2$$ \left(X^1\Sigma^+_g\right)$	&0.170066	&0.0837	&1.61890	&0.7416	&38268&1.9426&\cite{BJ12}	\\[1ex]
CO$\left(X^1\Sigma^+\right)$	&0.149936	&1.1392	&2.20481	&1.1283	&9.0540&2.2994&\cite{BJ12}\\[1ex]
ICl$\left(X^1\Sigma_g^{+}\right)$	&-0.086212	&4.55237	&2.008578	&2.3209	&17557&1.849159&\cite{BJ14}	\\[1ex]
ICl$\left(A^3\Pi_1\right)$	&-0.167208	&4.55237	&2.542557	&2.6850	&3814.7&2.178324&\cite{BJ14}\\[1ex]
ICl$\left(A'^3\Pi_2\right)$	&-0.157361	&4.55237	&2.373450	&2.6650	&4875&2.050745&\cite{BJ14}\\[1ex]
\hline\hline
\end{tabular}\label{Tab1}
\vspace*{-1pt}}
\end{table}
\begin{table}[!t]
{\scriptsize
\caption{ \small Comparison of the bound-state energy eigenvalues $-(E_{n\ell}-D)(eV)$ of H$_2$ and CO molecules for various $n$ and rotational $\ell$ quantum numbers in Tietz-Wei potential.} \vspace*{10pt}{
\begin{tabular}{ccccccccc}\hline\hline
{}&{}&{}&{}&{}&{}&{}&{}\\[-1.0ex]
\multicolumn{2}{c}{}&\multicolumn{4}{c}{H$_2$}&\multicolumn{3}{c}{CO}\\[1.5ex]
$n$&$\ell$	&PQR (Present)	&NU (Present)&NU \cite{BJ9}	&GPS \cite{AKR1} &PQR (Present)&NU (Present)	&NU\cite{BJ9}\\[1ex]\hline\hline
0	&0	&4.48160205	&4.48157183&4.4815718267&4.4815797825	&11.07378271&11.07370964	&11.07370964	\\[1ex]
	&5	&4.26706288	&4.26703273&4.2658220403&							&11.06668238&11.06660933	&11.06659606	\\[1ex]
	&10	&3.73741847	&3.73738873&3.7336304360&							&11.04775337&11.04768033	&11.04763173	\\[1ex]
5	&0	&2.28155518	&2.28153387&2.2669650930&2.2815913849	& 9.63231601& 9.63224796	&9.629868985	\\[1ex]
	&5	&2.12184477	&2.12182368&2.1070072990&							& 9.62560750& 9.62553946	&9.623148913	\\[1ex]
	&10	&1.72567170	&1.72565129&1.7105026080&							& 9.60772327& 9.60765524	&9.084920084	\\[1ex]
7	&0	&1.63001750	&1.62999964&1.6130911000&							& 9.08816748& 9.08810144	&9.084920084	\\[1ex]
	&5	&1.48916801 &1.48915041&1.4722799590&							& 9.08161259& 9.08154656	&9.078354525	\\[1ex]
	&10	&1.13910018	&1.13908334&1.1225356530&							& 9.06413799& 9.06407196	&9.060851440	\\[1ex]
\hline\hline
\end{tabular}\label{Tab2}}
\vspace*{-1pt}}
\end{table}
\begin{table}[!t]
{\scriptsize
\caption{\normalsize Bound-state energy eigenvalues for NO $\left(a^4\Pi_i\right)$, NO $\left(B^2\Pi_r\right)$, NO $\left(L'^2\phi\right)$, NO $\left(b^4\Sigma^{-}\right)$ and NO $(X^2\Pi_r)$ molecules for various $n$ and rotational $\ell$ quantum numbers in Tietz-Wei diatomic molecular potential. } \vspace*{10pt}{
\begin{tabular}{ccccccc}\hline\hline
{}&{}&{}&{}&{}&{}&{}\\[-1.0ex]
$n$&$\ell$&NO $\left(a^4\Pi_i\right)$&NO $\left(B^2\Pi_r\right)$&NO $\left(L'^2\phi\right)$&NO $\left(b^4\Sigma^{-}\right)$&NO $(X^2\Pi_r) $\\[2.5ex]\hline\hline
0	&0	&0.05724382	&-0.06511639	&-0.05748847	&-0.07561426	&-0.04775936	\\[1ex]
1	&0	&0.16924769	&-0.19664818	&-0.17491875	&-0.22970681	&-0.14364066	\\[1ex]
	&1	&0.16950472	&-0.19637388	&-0.17464813	&-0.22937970	&-0.14352509	\\[1ex]
2	&0	&0.27795077	&-0.33036890  &-0.29569047	&-0.38771825	&-0.24006757	\\[1ex]
	&1	&0.27820318	&-0.33009351	&-0.29541545	&-0.38738640	&-0.23995169	\\[1ex]
	&2	&0.27870799	&-0.32954274	&-0.29486542	&-0.38672273	&-0.23971993	\\[1ex]
3	&0	&0.38335537	&-0.46625489	&-0.41978250	&-0.54962300	&-0.33703799	\\[1ex]
	&1	&0.38360316	&-0.46597847	&-0.41950311	&-0.54928644	&-0.33692180	\\[1ex]
	&2	&0.38409872	&-0.46542563	&-0.41894434	&-0.54861334	&-0.33668942	\\[1ex]
	&3	&0.38484202	&-0.46459641	&-0.41810624	&-0.54760374	&-0.33634086	\\[1ex]
4	&0	&0.48546381	&-0.60428372	&-0.54717398	&-0.71539576	&-0.43454981	\\[1ex]
	&1	&0.48570698	&-0.60400632	&-0.54689024	&-0.71505453	&-0.43443332	\\[1ex]
	&2	&0.48619332	&-0.60345153	&-0.54632279	&-0.71437207	&-0.43420033	\\[1ex]
	&3	&0.48692277	&-0.60261938	&-0.54547165	&-0.71334844	&-0.43385085	\\[1ex]
	&4	&0.48789530	&-0.60150990	&-0.54433688	&-0.71198366	&-0.43338490	\\[1ex]
\hline\hline
\end{tabular}\label{Tab3}}
\vspace*{-1pt}}
\end{table}
\begin{table}
{\scriptsize
\caption{\normalsize Bound-state energy eigenvalues for H$_2$ $\left(X^1\Sigma^+_g\right)$, ICl $\left(X^1\Sigma_g^{+}\right)$, ICl $\left(A^3\Pi_1\right)$, ICl $\left(A'^3\Pi_2\right)$ and ICl $\left(B'O^{+}\right)$ molecules for various $n$ and rotational $\ell$ quantum numbers in Tietz-Wei diatomic molecular potential. } \vspace*{10pt}{
\begin{tabular}{ccccccc}\hline\hline
{}&{}&{}&{}&{}&{}&{}\\[-1.0ex]
$n$&$\ell$&H$_2$ $\left(X^1\Sigma^+_g\right)$&CO $\left(X^1\Sigma^+\right)$&ICl $\left(X^1\Sigma_g^{+}\right)$&ICl $\left(A^3\Pi_1\right)$&ICl $\left(A'^3\Pi_2\right)$\\[2.5ex]\hline\hline
0	&0	&0.26925518	&0.15070333	&-0.02388751	&-0.01317364	&-0.01400330	\\[1ex]
1	&0	&0.77877437	&0.44839050	&-0.07201091	&-0.03995431	&-0.04239940	\\[1ex]
	&1	&0.79255794	&0.44885861	&-0.07198238	&-0.03993286	&-0.04237766	\\[1ex]
2	&0	&1.25241802	&0.74134891	&-0.12061278	&-0.06734751	&-0.07134441	\\[1ex]
	&1	&1.26542046	&0.74181176	&-0.12058411	&-0.06732587	&-0.07132249	\\[1ex]
	&2	&1.29131118	&0.74273745	&-0.12052676	&-0.06728259	&-0.07127867	\\[1ex]
3	&0	&1.69133322	&1.02961057	&-0.16969191	&-0.09534658	&-0.10083355	\\[1ex]
	&1	&1.70358901	&1.03006820	&-0.16966309	&-0.09532475	&-0.10081146	\\[1ex]
	&2	&1.72799429	&1.03098343	&-0.16960544	&-0.09528108	&-0.10076729	\\[1ex]
	&3	&1.76433953	&1.03235625	&-0.16951898	&-0.09521559	&-0.10070103	\\[1ex]
4	&0	&2.09661354	&1.31320715	&-0.21924707	&-0.12394503	&-0.13086213	\\[1ex]
	&1	&2.10815488	&1.31365958	&-0.21921810	&-0.12392301	&-0.13083988	\\[1ex]
	&2	&2.13113862	&1.31456443	&-0.21916017	&-0.12387897	&-0.13079536	\\[1ex]
	&3	&2.16536983	&1.31592167	&-0.21907327	&-0.12381290	&-0.13072858	\\[1ex]
	&4	&2.21056327	&1.31773125	&-0.21895740	&-0.12372482	&-0.13063955	\\[1ex]
\hline\hline
\end{tabular}\label{Tab4}}
\vspace*{-1pt}}
\end{table}
\begin{equation}
\frac{1}{r^2}\approx\frac{1}{r_e^2}\left(D_0+D_1\frac{e^{-b_h(r-r_e)}}{1-c_he^{-b_h(r-r_e)}}+D_2\frac{e^{-2b_h(r-r_e)}}{\left(1-c_he^{-b_h(r-r_e)}\right)^2}\right),
\label{E8}
\end{equation}
with
\begin{subequations}
\begin{eqnarray}
D_0&=&1-\frac{1}{\alpha}(1-c_h)(3+c_h)+\frac{3}{\alpha^2}(1-c_h)^2, \ \ \ \ \ \ \lim_{c_h\rightarrow 0} D_0=1-\frac{3}{\alpha}+\frac{3}{\alpha^2}\\
D_1&=&\frac{2}{\alpha}(1-c_h)^2(2+c_h)-\frac{6}{\alpha^2}(1-c_h)^3, \ \ \ \ \ \ \lim_{c_h\rightarrow 0} D_1=\frac{4}{\alpha}-\frac{6}{\alpha^2}\\
D_2&=&-\frac{1}{\alpha}(1-c_h)^3(1+c_h)+\frac{3}{\alpha^2}(1-c_h)^4, \ \ \ \ \ \ \lim_{c_h\rightarrow 0} D_2=-\frac{1}{\alpha}+\frac{3}{\alpha^2},
\label{E9}
\end{eqnarray}
\end{subequations}
is a good approximation scheme to deal with the centrifugal potential term. Constant $\alpha=b_hr_e$ has been introduced for the sake of simplicity. Now, by inserting this approximation into equation (\ref{E7}) and then introducing a new transformation of the form $r\rightarrow\varrho=\frac{r-r_e}{r_e}$ through the mapping function $\varrho=f(r)$ with r in the domain $\left[\left.0, \infty\right.\right)$ or $\varrho$ in the domain $[-1, \infty]$, we obtain the following second order differential equation:
\begin{eqnarray}
&&\frac{1}{r_e^2}\frac{d^2R_{n\ell}(\varrho)}{d\varrho^2}+\frac{2\mu}{\hbar^2}\left[E_{n\ell}-V_{eff}(\varrho)\right]R_{n\ell}(\varrho)=0,\ \ \ \ \ \ \mbox{with}\\
&&V_{eff}(\varrho)=\left[\frac{\ell(\ell+1)D_0}{r_e^2}+\frac{2\mu D}{\hbar^2}+\frac{\frac{\ell(\ell+1)D_1}{r_e^2}+\frac{4\mu D}{\hbar^2}(c_h-1)}{e^{\alpha\varrho}-c_h}+\frac{\frac{\ell(\ell+1)D_2}{r_e^2}+\frac{2\mu D}{\hbar^2}(c_h-1)^2}{\left(e^{\alpha\varrho}-c_h\right)^2}\right]\frac{\hbar^2}{2\mu}.\nonumber
\label{E10}
\end{eqnarray}
The two  turning points are obtained by solving $V_{eff}(\varrho)-E_{n\ell}=0$ or $V_{eff}(\rho)-E_{n\ell}=0$ with $\rho=\left(e^{\alpha\varrho}-c_h\right)^{-1}$. Thus, it is easy to show that the turning points $\rho_a$ and $\rho_b$ are
\begin{subequations}
\begin{equation}
\rho_a=-\frac{\frac{\ell(\ell+1)D_1}{r_e^2}+\frac{4\mu D}{\hbar^2}(c_h-1)+\sqrt{\left[\frac{\ell(\ell+1)D_1}{r_e^2}+\frac{4\mu D}{\hbar^2}(c_h-1)\right]^2-4T_{n\ell}\left[\frac{\ell(\ell+1)D_2}{r_e^2}+\frac{2\mu D}{\hbar^2}(c_h-1)^2\right]}}{2T_{n\ell}}
\label{E11a}
\end{equation}
\begin{equation}
\rho_b=-\frac{\frac{\ell(\ell+1)D_1}{r_e^2}+\frac{4\mu D}{\hbar^2}(c_h-1)-\sqrt{\left[\frac{\ell(\ell+1)D_1}{r_e^2}+\frac{4\mu D}{\hbar^2}(c_h-1)\right]^2-4T_{n\ell}\left[\frac{\ell(\ell+1)D_2}{r_e^2}+\frac{2\mu D}{\hbar^2}(c_h-1)^2\right]}}{2T_{n\ell}}
\label{E11b}
\end{equation}
\end{subequations}
with the following sum and product properties:
\begin{equation}
\rho_a+\rho_b=-\frac{\ell(\ell+1)D_1{\hbar^2}+{4\mu r_e^2D}(c_h-1)}{T_{n\ell}r_e^2\hbar^2}\ \ \mbox{and}\ \ \rho_a\rho_b=\frac{\ell(\ell+1)D_2{\hbar^2}+{2\mu r_e^2D}(c_h-1)^2}{T_{n\ell}r_e^2\hbar^2}.
\label{E12}
\end{equation}
Furthermore, the momentum $k(\rho)$ between two turning points can be found as:
\begin{equation}
k(\rho)=\sqrt{\left[\frac{\ell(\ell+1)D_2}{r_e^2}+\frac{2\mu D}{\hbar^2}(c_h-1)^2\right]\left[(\rho_b-\rho)(\rho-\rho_a)\right]}.
\label{E13}
\end{equation}
The Riccati relation given by equation (\ref{E2}) can be re-written for the ground state as
\begin{equation}
-\frac{\alpha\rho(1+c_h\rho)}{r_e}\phi_0'(\rho)+\phi_0(\rho)^2=-\frac{2\mu}{\hbar^2}\left[E_{0\ell}-V_{eff}(\rho)\right].
\label{E14}
\end{equation}
Since the logarithmic derivative $\phi_0(\rho)$ for the ground state has one zero and no pole, it has to take the linear form in $\rho$. The only possible solution satisfying equation (\ref{E14}) is of the form $\phi_0(\rho)=\mathcal{A}+\mathcal{B}\rho$. The substitution of this expression into equation (\ref{E14}), one has the ground state energy eigenvalue 
\begin{eqnarray}
&&E_{0\ell}=\left[\frac{\ell(\ell+1)D_0}{r_e^2}-\mathcal{A}^2\right]\frac{\hbar^2}{2\mu}+D \ \ \ \mbox{with}\ \ \  \mathcal{A}=\frac{1}{2\mathcal{B}}\left[\frac{\ell(\ell+1)}{r_e^2}D_1+\frac{4\mu D}{\hbar^2}(c_h-1)\right]+\frac{\alpha}{2r_e}\nonumber\\
&&\mbox{and}\ \ \ \mathcal{B}=\frac{c_h\alpha}{2r_e}-\frac{c_h\alpha}{2r_e}\sqrt{1+\frac{4r_e^2}{\alpha^2c_h^2}\left[\frac{\ell(\ell+1)D_2}{r_e^2}+\frac{2\mu D}{\hbar^2}(c_h-1)^2\right]}
\label{E15}
\end{eqnarray}
We have now reached a position of calculating the integrals given by equation (\ref{E4}). The LHS integral can be calculated as follows:
\begin{eqnarray}
\int_{r_{A}}^{r_{B}}k(r)dr&=&r_e\int_{\varrho_{a}}^{\varrho_{b}}k(\varrho)d\varrho=-\frac{r_e}{\alpha}\int_{\rho_b}^{\rho_a}\frac{k(\rho)}{\rho(1+c_h\rho)}d\rho\nonumber\\
&=&-\frac{r_e}{\alpha}\sqrt{\frac{\ell(\ell+1)D_2}{r_e^2}+\frac{2\mu D}{\hbar^2}(c_h-1)^2}\int_{\rho_a}^{\rho_b}\frac{\sqrt{(\rho_b-\rho)(\rho-\rho_a)}}{\rho(1+c_h\rho)}d\rho\label{EE16}\\
&=&-\frac{\pi r_e}{\alpha}\sqrt{\frac{\ell(\ell+1)D_2}{r_e^2}+\frac{2\mu D}{\hbar^2}(c_h-1)^2}\left[\frac{\sqrt{(1+c_h\rho_a)(1+c_h\rho_b)}}{c_h}-\frac{1}{c_h}-\sqrt{\rho_a\rho_b}\right]\nonumber\\
&=&-\frac{\pi r_e}{\alpha}\left[\sqrt{T_{n\ell}-\frac{1}{c_h}\left[\frac{\ell(\ell+1)D_1}{r_e^2}+\frac{4\mu D}{\hbar^2}(c_h-1)\right]+\frac{R_{T}}{c_h^2}}-\frac{\sqrt{R_{T}}}{c_h}-\sqrt{T_{n\ell}}\right]\nonumber\\
&&\mbox{with}\ \ \ T_{n\ell}=\left[\frac{\ell(\ell+1)D_0}{r_e^2}+\frac{2\mu}{\hbar^2}(D-E_{n\ell})\right]\ \ \ \mbox{and}\ \ \ R_{T}=\frac{\ell(\ell+1)D_2\hbar^2}{2\mu r_e^2}+D(c_h-1)^2,\nonumber
\end{eqnarray}
where we have utilized the properties (\ref{E12}) and the integral relation given by 
\begin{eqnarray}
\int_{x_a}^{x_b}\frac{\sqrt{(x_b-x)(x-x_a)}}{x(1+Qx)}dx=\pi\left[\frac{\sqrt{(Qx_a+1)(Qx_b+1)}}{Q}-\frac{1}{Q}-\sqrt{x_ax_b}\right].
\label{Appendix1}
\end{eqnarray}
Now, simply by replacing $E_{n\ell}$ in the above equation (\ref{EE16}) by $E_{0\ell}$ given by equation (\ref{E15}), and $T_{n\ell}$ as $T_{0\ell}=\left[\frac{\ell(\ell+1)D_0}{r_e^2}+\frac{2\mu}{\hbar^2}\left(D-E_{0\ell}\right)\right]$, we obtain the integral in the RHS of equation (\ref{E4}) as
\begin{eqnarray}
\int_{r_{0a}}^{r_{0b}}k_0(r)dr&=&-\frac{\pi r_e}{\alpha}\left[\sqrt{T_{0\ell}-\frac{1}{c_h}\left[\frac{\ell(\ell+1)D_1}{{r_e^2}}+\frac{4\mu D}{\hbar^2}(c_h-1)\right]+\frac{R_{T}}{c_h^2}}-\frac{\sqrt{R_{T}}}{c_h}-\sqrt{T_{0\ell}}\right]\nonumber\\
&=&\frac{\pi r_e}{\alpha c_h}\left[\sqrt{R_{T}}+\mathcal{B}\right].
\label{E17}
\end{eqnarray}
With equations (\ref{EE16}), (\ref{E17}) and (\ref{E4}), we can deduce the following relation
\begin{eqnarray}
&&-\frac{\pi r_e}{\alpha}\left[\sqrt{T_{n\ell}-\frac{1}{c_h}\left[\frac{\ell(\ell+1)D_1}{{r_e^2}}+\frac{4\mu D}{\hbar^2}(c_h-1)\right]+\frac{R_{T}}{c_h^2}}-\sqrt{T_{n\ell}}\right]-\frac{\pi r_e}{\alpha c_h}\mathcal{B}=n\pi.\nonumber\\
&&\sqrt{T_{n\ell}-\frac{1}{c_h}\left[\frac{\ell(\ell+1)D_1}{{r_e^2}}+\frac{4\mu D}{\hbar^2}(c_h-1)\right]+\frac{R_{T}}{c_h^2}}=-\left(n+\frac{r_e}{\alpha c_h}\mathcal{B}\right)\frac{\alpha}{r_e}+\sqrt{T_{n\ell}}
\label{E18}
\end{eqnarray}
On squaring up both sides of equation (\ref{E18}), it is straightforward to show that the energy eigenvalues equation can be found as
\begin{eqnarray}
E_{n\ell}&=&\frac{\hbar^2\ell(\ell+1)D_0}{2\mu r_e^2}+D-\frac{\alpha^2\hbar^2}{2\mu r_e^2}\left[\frac{\eta^2+\frac{\ell(\ell+1)}{\alpha^2c_h^2}(D_1c_h-D_2)+\frac{2\mu Dr_e^2}{\alpha^2\hbar^2}\left(1-\frac{1}{c_h^2}\right)}{2\eta}\right]^2\\
&&\mbox{with}\ \ \eta=n+\frac{1}{2}+\frac{1}{2}\sqrt{1+\frac{4}{c_h^2}\left(\frac{D_2\ell(\ell+1)}{\alpha^2}+\frac{2\mu Dr_e^2}{\alpha^2\hbar^2}(1-c_h)^2\right)}.\nonumber
\label{E19}
\end{eqnarray}
\subsection{Shifted Deng-Fan molecular potential}
The shifted Deng-Fan molecular potential model we examine in this section is defined as  \cite{SM15,BJ12}
\begin{equation}
V(r)=D\left(1-\frac{b}{e^{\beta r}-1}\right)^2-\bar{D},\ \ \ \ \ \ b = e^{ar_{e}} - 1,
\label{E20}
\end{equation}
where $(D,\bar{D})$, $b$ and $\beta$ are three parameters representing the dissociation energy, the position of the minimum $r_e$ and the range of the potential respectively. Very recently,  Wang and co-workers found that the  Manning-Rosen, Deng-Fan and Schi\"{o}berg potential are not better than the traditional Morse potential in simulating the atomic interaction for diatomic molecules \cite{BJ13}. In order to overcome this problem, Hamzavi et al. suggested a modification to the Deng-Fan potential, which they referred to as the shifted Deng-Fan potential (sDF) \cite{SM15}. This modification is simply a Deng-Fan potential \cite{B6,B7} shifted by dissociation energy ${D}$ \cite{SM15}. The researchers \cite{SM15} examined the Schr\"{o}dinger equation with this potential and applied their results to some diatomic molecules \cite{SM15}. From their plot for the shifted Deng-Fan potential and the Morse potential using the parameters set for $H_{2}$ diatomic molecule, it was shown that the two potentials are very close to each other for large values of $r$ in the regions $r\approx r_{e}$ and $r>r_{e}$, but they are very different at $ r \approx 0$. Also, if both the Deng-Fan and the shifted Deng-Fan potentials are deep (that is,  $D>>1$) they could be well approximated by a harmonic oscillator in the region $r\approx r_{e}$ \cite{SM15}.

\begin{figure}[!htb]
\centering\includegraphics[height=85mm,width=150mm]{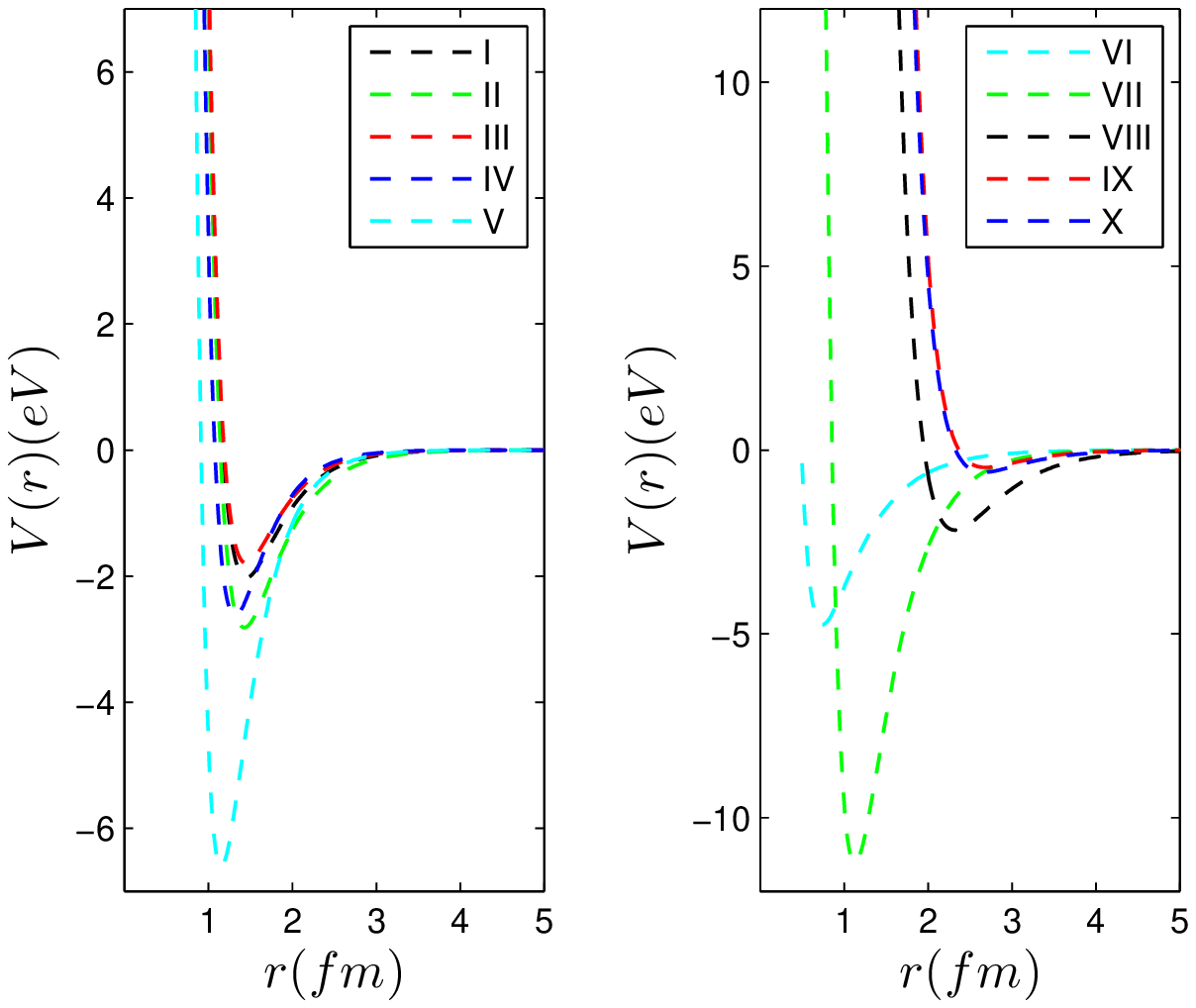}
\caption{{\protect\footnotesize Shape of shifted Deng-Fan diatomic molecular potential for different diatomic molecules:  (I) NO$\left(a^4\Pi_i\right)$ (II) NO$\left(B^2\Pi_r\right)$ (III) NO$\left(L'^2\phi\right)$ (IV) NO$\left(b^4\Sigma^{-}\right)$ (V) NO$(X^2\Pi_r)$ (VI) H$_2\left(X^1\Sigma^+_g\right)$ (VII) CO$\left(X^1\Sigma^+\right)$ (VIII) ICl$\left(X^1\Sigma_g^{+}\right)$ (IX) ICl$\left(A^3\Pi_1\right)$ (X) ICl$\left(A'^3\Pi_2\right)$.}}
\label{fig7}
\end{figure}

In Figure (\ref{fig7}), we study the variation of this potential with respect to some diatomic molecules of interest given in table \ref{Tab1}. Now inserting this potential into the Schr\"{o}dinger equation, and then use the approximation of the form \cite{BJ12}:
\begin{equation}
\frac{1}{r^2}=\left[d_0+\frac{e^{-\beta r}}{\left(1-e^{-\beta r}\right)^2}\right],
\label{E21}
\end{equation}

\begin{figure}[!t]
 \includegraphics[height=85mm,width=150mm]{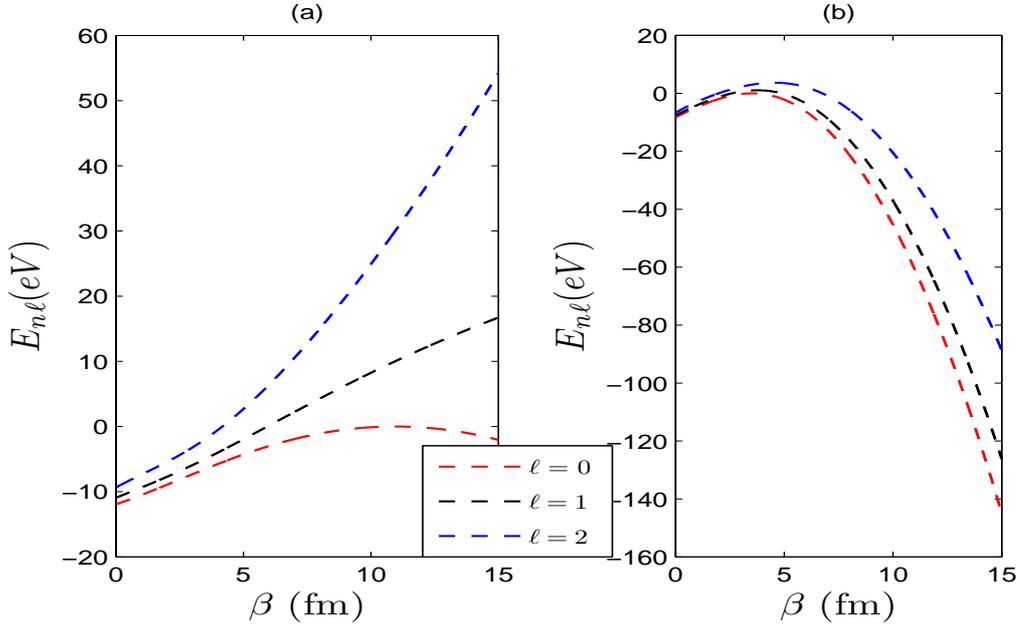}
\caption{{\protect\small (a) The variation of the ground state energy state for various values of $\ell$ as a function of the potential range $\beta$. We choose $\mu=1$,  $r_e=0.8$ and $D=15$. (b) The variation of the first excited energy state for various $\ell$ as a function of the potential range $\beta$.}}
\label{fig8}
\end{figure}

\begin{figure}[!t]
 \includegraphics[height=85mm,width=150mm]{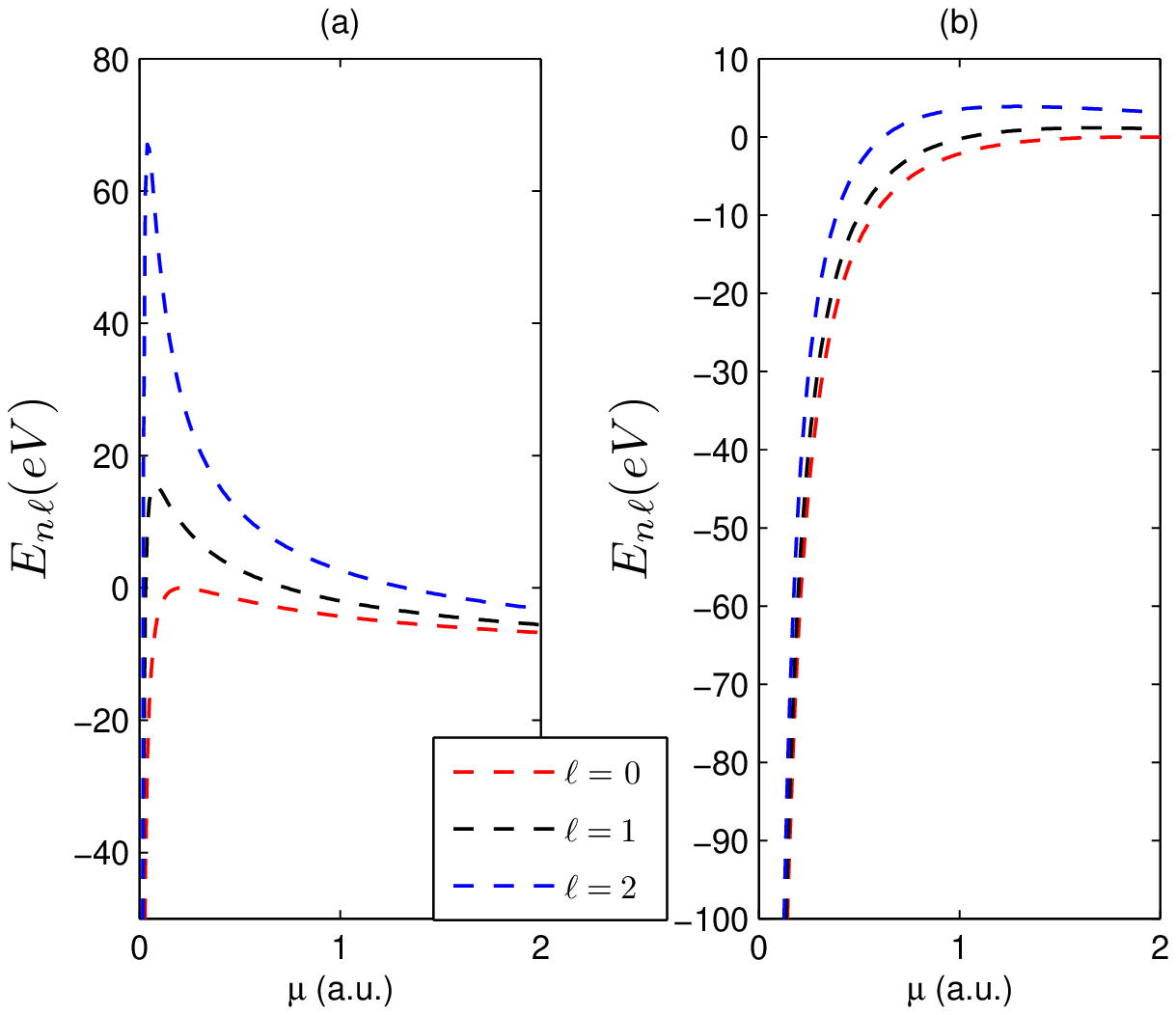}
\caption{{\protect\small (a) The variation of the ground state energy state for various values of $\ell$ as a function of the particle mass $\mu$. We choose $\beta=5$,  $r_e=0.8$ and $D=15$. (b) The variation of the first excited energy state for various $\ell$ as a function of the particle mass $\mu$.}}
\label{fig9}
\end{figure}

\begin{figure}[!t]
\includegraphics[height=85mm,width=150mm]{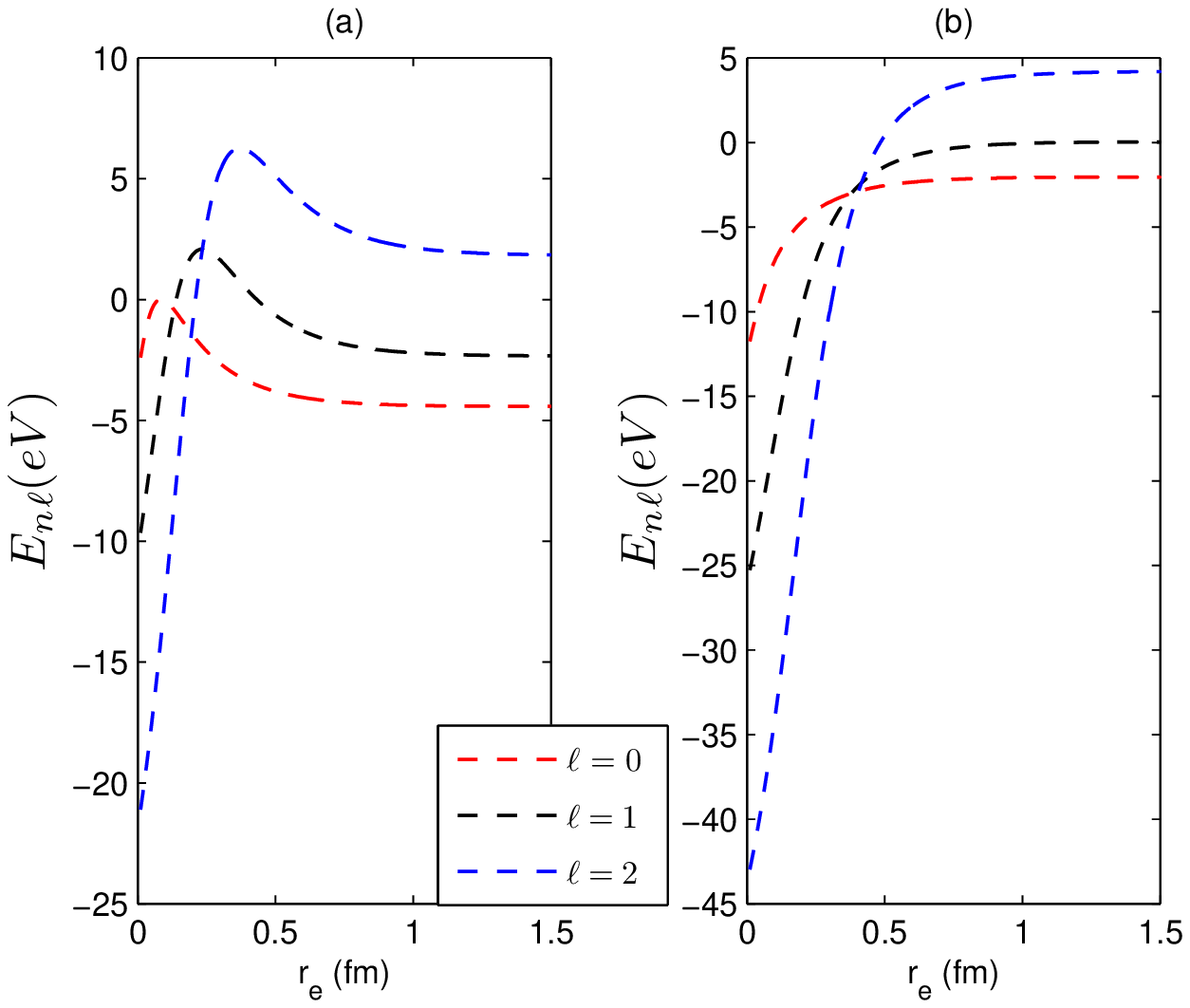}
\caption{{\protect\small (a) The variation of the ground state energy state for various values of $\ell$ as a function of the molecular bond length $r_e$. We choose $\mu=1$, $\beta=5$ and $D=15$. (b) The variation of the first excited energy state for various $\ell$ as a function of the molecular bond length $r_e$}}
\label{fig10}
\end{figure}

\begin{figure}[!t]
\includegraphics[height=85mm,width=150mm]{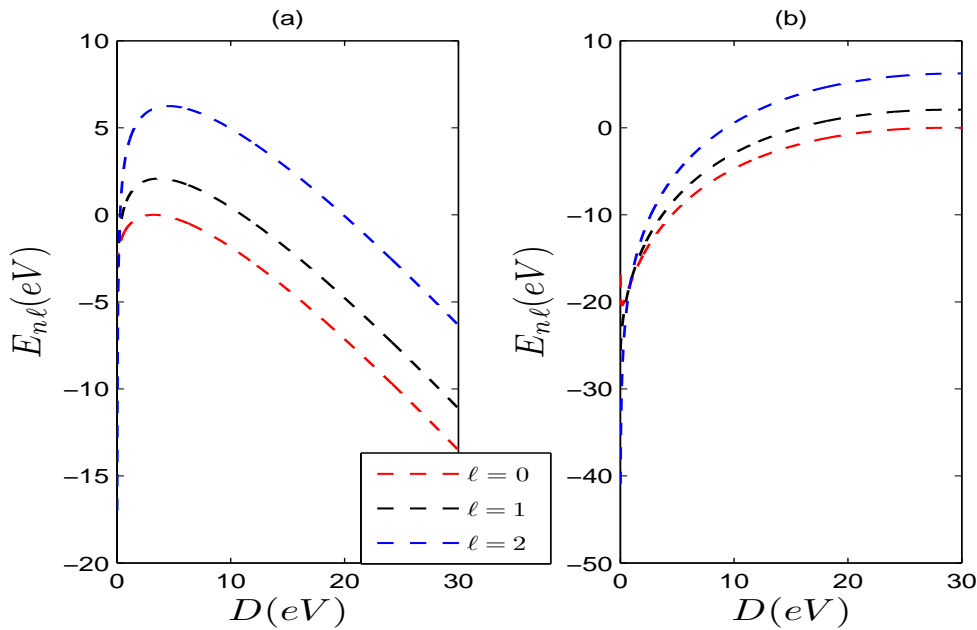}
\caption{{\protect\small (a) The variation of the ground state energy state for various values of $\ell$ as a function of the dissociation energy $D_e$. We choose $\mu=1$,  $\beta=5$ and $r_e=0.8$. (b) The variation of the first excited energy state for various $\ell$ as a function of the dissociation energy $D$}}
\label{fig11}
\end{figure}
\begin{table}[!t]
{\scriptsize
\caption{ \small Comparison of the bound-state energy eigenvalues $-E_{n\ell}(eV)$ of H$_2$ and CO molecules for various $n$ and rotational $\ell$ quantum numbers in sDF diatomic molecular potential.} \vspace*{10pt}{
\begin{tabular}{ccccccccccc}\hline\hline
\multicolumn{2}{c}{}& \multicolumn{3}{c}{H$_2$ Diatomic Molecule}&&\multicolumn{2}{c}{}&\multicolumn{3}{c}{CO Diatomic Molecule}\\[1.5ex]
{}&{}&{}&{}&{}&{}&{}&{}&{}&{}&{}\\[-1.0ex]
$n$&$\ell$&Present  &GPS \cite{RO1} &AIM \cite{BJ12}&N-U \cite{SM15}&&Present&GPS \cite{RO1} &AIM \cite{BJ12}&N-U\cite{BJ9} (CO)\\[1ex]\hline\hline
0	&0	&4.39461978& 4.39462330967	&4.394619779	&4.39444	&&11.08074990& 11.0807513815	&11.08075178	&11.08068	\\[1ex]
	&5	&4.17661316&								&4.176618048	&4.17644	&&11.07253746&						    &11.07253985	&11.07247	\\[1ex]
	&10	&3.62182049&								&3.621838424	&3.62165	&&11.05064208&						    &11.05064581	&11.05057	\\[1ex]
5	&0	&1.75845157&								&1.758451567	&1.75835	&& 9.68814442&						    &9.688146187	&9.68809	\\[1ex]
	&5	&1.61740572&								&1.617410615	&1.61731	&& 9.68022402&						    &9.680226284	&9.68017	\\[1ex]
	&10	&1.26043371&								&1.260451640	&1.26034	&& 9.65910731&						    &9.659110919	&9.65905	\\[1ex]
7	&0	&1.07763699&								&1.077636993	&1.07756	&& 9.15916229&						    &9.159164003	&9.15911	\\[1ex]
	&5	&0.96180989&								&0.961814782	&0.96174	&& 9.15135744&						    &9.151359661	&9.15131	\\[1ex]
	&10	&0.66982613&								&0.669844065	&0.66976	&& 9.13054886&						    &9.130552425	&9.13050	\\[1ex]
\hline\hline
\end{tabular}\label{Tab5}}
\vspace*{-1pt}}
\end{table}
\begin{table}[!t]
{\scriptsize
\caption{\normalsize Bound-state energy eigenvalues for NO $\left(a^4\Pi_i\right)$, $NO \left(B^2\Pi_r\right)$, NO $\left(L'^2\phi\right)$, NO $\left(b^4\Sigma^{-}\right)$ and NO $(X^2\Pi_r)$ molecules for various $n$ and rotational $\ell$ quantum numbers in sDF diatomic molecular potential. } \vspace*{10pt}{
\begin{tabular}{ccccccc}\hline\hline
{}&{}&{}&{}&{}&{}&{}\\[-1.0ex]
$n$&$\ell$&NO $\left(a^4\Pi_i\right)$&NO $\left(B^2\Pi_r\right)$&NO $\left(L'^2\phi\right)$&NO $\left(b^4\Sigma^{-}\right)$&NO $(X^2\Pi_r) $\\[2.5ex]\hline\hline
0	&0	&-1.96954814	&-2.75045354	&-1.73990613	&-2.54988834	&-6.49084320	\\[1ex]
1	&0	&-1.85431702	&-2.61950265	&-1.62684906	&-2.40042016	&-6.24919850	\\[1ex]
	&1	&-1.85394591	&-2.61914169	&-1.62645839	&-2.39995170	&-6.24866000	\\[1ex]
2	&0	&-1.74265661	&-2.49186027	&-1.51767759	&-2.25557591	&-6.01232867	\\[1ex]
	&1	&-1.74228862	&-2.49150219	&-1.51729002	&-2.25511093	&-6.01179390	\\[1ex]
	&2	&-1.74155263	&-2.49078605	&-1.51651490	&-2.25418099	&-6.01072438	\\[1ex]
3	&0	&-1.63455757	&-2.36751730	&-1.41238208	&-2.11534494	&-5.78022207	\\[1ex]
	&1	&-1.63419268	&-2.36716213	&-1.41199768	&-2.11488343	&-5.77969103	\\[1ex]
	&2	&-1.63346278	&-2.36645179	&-1.41122876	&-2.11396043	&-5.77862896	\\[1ex]
	&3	&-1.63236803	&-2.36538629	&-1.41007542	&-2.11257593	&-5.77703588	\\[1ex]
4	&0	&-1.53001045	&-2.24646485	&-1.31095323	&-1.97971663	&-5.55286713	\\[1ex]
	&1	&-1.52964860	&-2.24611256	&-1.31057185	&-1.97925858	&-5.55233981	\\[1ex]
	&2	&-1.52892485	&-2.24540801	&-1.30980918	&-1.97834250	&-5.55128517	\\[1ex]
	&3	&-1.52783925	&-2.24435119	&-1.30866514	&-1.97696839	&-5.54970324	\\[1ex]
	&4	&-1.52639185	&-2.24294214	&-1.30713977	&-1.97513628	&-5.54759404	\\[1ex]
\hline\hline
\end{tabular}\label{Tab6}}
\vspace*{-1pt}}
\end{table}
\begin{table}
{\scriptsize
\caption{\normalsize Bound-state energy eigenvalues for H$_2$ $\left(X^1\Sigma^+_g\right)$, ICl $\left(X^1\Sigma_g^{+}\right)$, ICl $\left(A^3\Pi_1\right)$, ICl $\left(A'^3\Pi_2\right)$ and ICl $\left(B'O^{+}\right)$ molecules for various $n$ and rotational $\ell$ quantum numbers in sDF diatomic molecular potential. } \vspace*{10pt}{
\begin{tabular}{ccccccc}\hline\hline
{}&{}&{}&{}&{}&{}&{}\\[-1.0ex]
$n$&$\ell$&H$_2$ $\left(X^1\Sigma^+_g\right)$&CO $\left(X^1\Sigma^+\right)$&ICl $\left(X^1\Sigma_g^{+}\right)$&ICl $\left(A^3\Pi_1\right)$&ICl $\left(A'^3\Pi_2\right)$\\[2.5ex]\hline\hline
0	&0	&-4.40110165	&-11.08006149	&-2.152719356	&-0.459938084	&-0.59052741	\\[1ex]
1	&0	&-3.75459356	&-10.79430704	&-2.104958026	&-0.434430900	&-0.56321617	\\[1ex]
	&1	&-3.74109390	&-10.79376640	&-2.104907368	&-0.434368596	&-0.56316009	\\[1ex]
2	&0	&-3.17123615	&-10.51257523	&-2.057739707	&-0.409653282	&-0.53655412	\\[1ex]
	&1	&-3.15882350	&-10.51203847	&-2.057689131	&-0.409591035	&-0.53649810	\\[1ex]
	&2	&-3.13406357	&-10.51096496	&-2.057587980	&-0.409466541	&-0.53638607	\\[1ex]
3	&0	&-2.64736026	&-10.23485368	&-2.011064148	&-0.385605052	&-0.51054106	\\[1ex]
	&1	&-2.63597228	&-10.23432079	&-2.011013655	&-0.385542864	&-0.51048511	\\[1ex]
	&2	&-2.61325755	&-10.23325500	&-2.010912669	&-0.385418487	&-0.51037321	\\[1ex]
	&3	&-2.57933793	&-10.23165635	&-2.010761189	&-0.385231921	&-0.51020535	\\[1ex]
4	&0	&-2.17959586	& -9.96113006	&-1.964931100	&-0.362286035	&-0.48517681	\\[1ex]
	&1	&-2.16917525	& -9.96060101	&-1.964880689	&-0.362223905	&-0.48512092	\\[1ex]
	&2	&-2.14839142	& -9.95954293	&-1.964779867	&-0.362099645	&-0.48500914	\\[1ex]
	&3	&-2.11735861	& -9.95795584	&-1.964628634	&-0.361913255	&-0.48484148	\\[1ex]
	&4	&-2.07624689	& -9.95583977	&-1.964426991	&-0.361664734	&-0.48461792	\\[1ex]
\hline\hline
\end{tabular}\label{Tab7}}
\vspace*{-1pt}}
\end{table}
the effective potential takes the following form:
\begin{eqnarray}
&&V_{eff}(y)=P+Qy+Ry^2 \ \ \ \ \mbox{with}\ \ \ P=D-\bar{D}+\frac{\ell(\ell+1)\beta^2d_0\hbar^2}{2\mu}\\
&&Q=\frac{\ell(\ell+1)\beta^2\hbar^2}{2\mu}-2Db \ \ \ \mbox{and}\ \ \ R=\frac{\ell(\ell+1)\beta^2d_0\hbar^2}{2\mu}+Db^2,\nonumber
\label{E22}
\end{eqnarray}
after an appropriate coordinate transformation of the form $y=\left(e^{\beta r}-1\right)^{-1}$ has been introduced. Now, we can write the non-linear Riccati equation for the ground state as
\begin{equation}
-ay(1+y)\phi_0'(y)+\phi_0^2(y)=-\frac{2\mu}{\hbar^2}\left[E_{0\ell}-V_{eff}(y)\right]
\label{E23}
\end{equation}
Since the logarithmic derivative $\phi_0(y)$ for the ground state has one zero and no pole, it has to take the linear form in $y$. Thus, we assume the following solution for the ground state
\begin{equation}
\phi_0(y)=A+By.
\label{E24}
\end{equation}
By putting equation (\ref{E24}) into (\ref{E23}) and then solve the non-linear Riccati equation, it is straightforward to obtain the ground state energy and values of A and B as
\begin{equation}
E_{0\ell}=P-\frac{\hbar^2A^2}{2\mu}\ \ \ \mbox{with}\ \ \ A=\frac{\mu}{\hbar^2}\frac{Q-R}{B}+\frac{B}{2} \ \ \mbox{and}\ \ \ B=\frac{\beta}{2}+\frac{1}{2}\sqrt{\beta^2+\frac{8\mu R}{\hbar^2}}.
\label{E25}
\end{equation}
Furthermore, in a similar fashion to the previous problem, the two turning points as well as their sum and product properties are given by
\begin{eqnarray}
&&y_a=-\frac{Q}{2R}-\frac{1}{2R}\sqrt{Q^2-4R(P-E_{n\ell})},\ \ \ \mbox{and}\ \ \ y_b=-\frac{Q}{2R}+\frac{1}{2R}\sqrt{Q^2-4R(P-E_{n\ell})}\nonumber\\
&&y_a+y_b=-\frac{Q}{R},\ \ \ \ y_ay_b=\frac{P-E_{n\ell}}{R}  \ \ \ \mbox{and} \ \ \  k(y)=\sqrt{\frac{2\mu R}{\hbar^2}}\left[-\left(y-y_a\right)\left(y-y_b\right)\right]^{1/2}.
\label{E26}
\end{eqnarray}
Now, we have all necessary tools required to perform our calculations. Therefore, we proceed to calculate integral (\ref{E4})
\begin{eqnarray}
\int_{r_{a}}^{r_{b}}k(r)dr&=&-\int_{y_{a}}^{y_{b}}\frac{k(y)}{\beta y(1+y)}dy=-\int_{y_{a}}^{y_{b}}\sqrt{\frac{2\mu R}{\beta^2\hbar^2}}\frac{\left[\left(y-y_a\right)\left(y_b-y\right)\right]^{1/2}}{y(1+y)}dy\nonumber\\
&=&-\frac{\pi}{\beta}\sqrt{\frac{2\mu R}{\hbar^2}}\left[\sqrt{(1+y_a)(1+y_b)}-1-\sqrt{y_ay_b}\right]\\
&=&-\frac{\pi}{\beta}\sqrt{\frac{2\mu R}{\hbar^2}}\left[\frac{\sqrt{R-Q+P-E_{n\ell}}}{R}-1-\sqrt{\frac{P-E_{n\ell}}{R}}\right],\nonumber
\label{E27}
\end{eqnarray}
where we have used the following standard integral
\begin{equation}
\int_{y_{a}}^{y_{b}}\frac{\left[-\left(y-y_a\right)\left(y-y_b\right)\right]^{1/2}}{y(1+y)}=\pi\left[\sqrt{(y_a+1)(y_b+1)}-1-\sqrt{y_ay_b}\right].
\label{E28}
\end{equation}
Furthermore, we can find 
\begin{eqnarray}
\int_{r_{0a}}^{r_{0b}}k_0(r)dr&=&-\int_{y_{0a}}^{y_{0b}}\frac{k(y)}{\beta y(1+y)}dy=-\int_{y_{0a}}^{y_{0b}}\sqrt{\frac{2\mu R}{a^2\hbar^2}}\frac{\left[-\left(y-y_a\right)\left(y-y_b\right)\right]^{1/2}}{y(1+y)}\nonumber\\
&=&-\frac{\pi}{\beta}\sqrt{\frac{2\mu R}{\hbar^2}}\left[\sqrt{(1+y_a)(1+y_b)}-1-\sqrt{y_ay_b}\right]\\
&=&-\frac{\pi}{\beta}\sqrt{\frac{2\mu R}{\hbar^2}}\left[\frac{\sqrt{R-Q+P-E_{0\ell}}}{R}-1-\sqrt{\frac{P-E_{0\ell}}{R}}\right]\nonumber\\
&=&-\frac{\pi}{\beta}\sqrt{\frac{2\mu R}{\hbar^2}}\left[\frac{\hbar^2}{2\mu R}(A-B)-\sqrt{\frac{P-E_{0\ell}}{R}}-1\right]=\frac{\pi}{\beta}\sqrt{\frac{2\mu R}{\hbar^2}}\left[B\sqrt{\frac{\hbar^2}{2\mu R}}+1\right].\nonumber
\label{E29}
\end{eqnarray}
From equations (\ref{E4}), (\ref{E27}) and (\ref{E29}), we can find the energy spectrum for the sDF as
\begin{equation}
E_{n\ell}=D(b+1)^2+\frac{\ell(\ell+1)\beta^2\hbar^2d_0}{2\mu}-\frac{\hbar^2\beta^2}{2\mu}\left[\frac{\left(\frac{B}{\beta}+n\right)}{2}+\frac{2\mu Db(b+2)}{2\hbar^2\beta^2\left(\frac{B}{\beta}+n\right)}\right]-\bar{D}.
\label{E30}
\end{equation}
\section{Calculation of the eigenfunctions}
Eigenfunctions-eigenvalue relation is very important in quantum mechanics because of its prominence in the equations which relate the mathematical formalism of the theory with physical results. eigenfunctions could be considered as trial functions in variational-type procedures for deriving energy levels anl also for computing line intensities. Since proper quantization rule cannot be used to obtain these eigenfunctions, we therefore resort to using the recently proposed formula method \cite{NEQ1}. This method is very easy to use in obtaining not only the eigenfunctions but also energy eigenvalues. In the approach, it is required  to transform the Schr\"{o}dinger equation with two solvable quantum molecular systems-Tietz-Wei and shifted Deng-Fan potential models into the form given by equation (1) of ref. \cite{NEQ1} via an apprropriate coordinate transformation of the form $\tau=e^{\beta\varrho}$ (for TH) and $t=e^{-\alpha r}$ (for sDF), which maintained the finiteness of the transformed wave functions on the boundary conditions to have
\begin{subequations}
\begin{eqnarray}
&&\frac{d^2R_{n\ell}(\tau)}{d\tau^2}+\frac{1}{\tau}\frac{dR_{n\ell}(\tau)}{d\tau}+\frac{1}{\tau^2(1-c_h\tau)^2}\left\{\left[\frac{2\mu r_e^2}{\hbar^2\alpha^2}(E_{n\ell}-D)-\frac{\ell(\ell+1)}{\alpha^2}D_0\right]\right.\nonumber\\
&&\left.+\left[-2c_h\left(\frac{2\mu r_e^2E_{n\ell}}{\alpha^2\hbar^2}-\frac{\ell(\ell+1)}{\alpha^2}D_0\right)+\frac{4\mu r_e^2D}{\hbar^2\alpha^2}-\frac{\ell(\ell+1)}{\alpha^2}D_1\right]\tau\right.\nonumber\\
&&\left.+\left[c_h^2\left(\frac{2\mu r_e^2E_{n\ell}}{\alpha^2\hbar^2}-\frac{\ell(\ell+1)}{\alpha^2}D_0\right)+\frac{\ell(\ell+1)}{\alpha^2}\left(D_1c_h-D_2\right)-\frac{2\mu r_e^2D}{\hbar^2\alpha^2}\right]\tau^2\right\}R_{n\ell}(\tau)=0,\nonumber\\
\label{E31a}
\end{eqnarray}
\begin{eqnarray}
&&\frac{d^2R_{n\ell}(t)}{dt^2}+\frac{1}{t}\frac{dR_{n\ell}(t)}{dt}+\frac{1}{t^2(1-t)^2}\left[\frac{2\mu}{\hbar^2}\left(E_{n\ell}-D\right)-\frac{2\mu Db}{\beta^2\hbar^2}(b+2)-\ell(\ell+1)d_0+t\left(\frac{4\mu bD}{\beta^2\hbar^2}\right.\right.\nonumber\\
&&\left.\left.-\frac{4\mu}{\beta^2\hbar^2}\left(E_{n\ell}-D\right)-\ell(\ell+1)(1-2d_0)\right)+t^2\left(\frac{2\mu}{\beta^2\hbar^2}\left(E_{n\ell}-D\right)-\ell(\ell+1)d_0\right)\right]R_{n\ell}(t)=0.\nonumber\\
\label{E31b}
\end{eqnarray}
\end{subequations}
Considering equation (\ref{E31a}) with reference to \cite{NEQ1}, $k_1$, $k_2$, $k_3$, $A_{T-H}$, $B_{T-H}$ and $C_{T-H}$ can be found. Then, parameters $k_4$ and $k_5$ can be obtained as
\begin{equation}
k_4=\sqrt{\left[\frac{2\mu r_e^2}{\hbar^2\alpha^2}(D-E_{n\ell})+\frac{\ell(\ell+1)}{\alpha^2}D_0\right]}\ \ \mbox{and} \ \ k_5=\frac{1}{2}\left\{1+\sqrt{1+\frac{4}{c_h^2}\left[\frac{\ell(\ell+1)}{\alpha^2}D_2+\frac{2\mu r_e^2D}{\hbar^2\alpha^2}(1-c_h)^2\right]}\right\},
\label{E32}
\end{equation}
Hence, the eigenfunctions for TW can be found as
 \begin{eqnarray}
R_{n\ell}(\varrho)&=&N_{n\ell}e^{-k_4\alpha\varrho}(1-c_he^{-\alpha\varrho})^{k_5}\ _2F_1\left(-n, n+2(k_4+k_5); 2k_4+1, c_he^{-\alpha\varrho}\right).
\label{E33}
\end{eqnarray}
Similarly, the eigenfunctions for sDF can be found as
\begin{equation}
R_{n\ell}(z)=N_{n\ell}t^w(1-t)^{v}\ _2F_1(-n, 2\left(w+v\right); 2w+1; t),
\label{E23}
\end{equation}
where
\begin{equation}
w=\sqrt{-\left(\frac{2\mu}{\hbar^2}\left(E_{n\ell}-D\right)-\frac{2\mu Db}{\beta^2\hbar^2}(b+2)+\ell(\ell+1)d_0\right)} \ \ \ \mbox{and}\ \ v=\frac{1}{2}+\sqrt{\left(\ell+\frac{1}{2}\right)+\frac{2\mu Db^2}{\beta^2\hbar^2}}.
\end{equation}
\section{Numerical Results and Discussion}
In Figure \ref{fig1}, we plotted the Tietz-Wei (TW) potential for different diatomic molecules. In what follows, to see the behavior of the ground $n=0$ and first excited $n=1$ states, we plotted the energy for these states with potential parameters for three different orbital states $\ell=0,1,2$. In Figure \ref{fig2}, we show the variation of $E_{n,\ell}$ with the potential constant $c_h$. It shows that for $c_h < 0$, the energy is negative whereas when $c_h > 0$, the energy is positive for $n=0$. On the other hand, for $n=1$, the energy becomes strongly bound for $c_h < 0$ and moves toward the negative energy for $c_h > 0$. The   $c_h = 0$ represents the Morse energy. The best choice $c_h = 0.03$ restores the results of Morse potential. At this value the energy curves coincide and have same behavior for $\ell=0,1,2$. Figure \ref{fig3} shows the variation of $E_{n,\ell}$ with the reduced mass $\mu$ for three orbital states. The energy is very similar for $0.1 <\mu < 1.0$ when $n=1$ but different when $n=0$. Its seen that when $\mu$ increases for more than 0.3, the energy spectrum becoming positive for ground state while for excited state, it is positive for any value of $\mu$ 

In Figure \ref{fig4}, we plotted the variation of $E_{n,\ell}$ with the potential parameter $b_h$. It is increasing in the positive direction within the interval $0 < b_h < 8$ when $n=0$. However, when $n=1$, the energy increases in positive side for $0< b_h <4$ and increases in the negative side for $4< b_h < 8$.  In Figure \ref{fig5} we show the variation of the energy states $E_{n,\ell}$ as a function of molecular bond length $r_e$. The ground state energy drops with nearly 2eV  for all orbital states  at $r_e=0.8fm$ and $r_e=0.9fm$ whereas the first excited state has drop of about 0.45eV and coincide  at 0.56fm.  Finally, Figure \ref{fig6} demonstrates the energy versus the well depth $D$. Its seen that the ground state energy span from negative to positive spectrum at $D = 2eV$. for orbital states $\ell=0, 1, 2$.  However the first excited energy state span from negative to positive spectrum at $D=8eV$ for all orbital states.

A very similar behavior to TW potential model (sDF shape) for various molecules is shown in Figure 7. In addition we have obtained the energy spectrum for different diatomic molecules with the help of TW molecular model for various states using the model potential parameters in Table \ref{Tab1}. This spectroscopic parameter are taken from Refs. \cite{BJ12}, \cite{BJ14} and \cite{BJ15} and the conversion factors used are taken from NIST database \cite{BJ16}: $1cm^{-1}=1.239841930$eV, $\hbar c=1973.29eV\AA$ and $1amu=931.494061Mev/c^2$. In Table \ref{Tab2}, we test the accuracy of the method utilized in this study by finding the energy spectra of H$_2$ and CO diatomic molecules. We found that the spectrum obtained by NU method in \cite{BJ9} have some error in the Maple codes. We therefore re-compute these spectrum in the present work for the sake of comparison. As it can been seen from the table, our results are very close to the ones of the Nikiforov-Uvarov method. Tables \ref{Tab3} and \ref{Tab4} present the spectrum  for H$_2 \left(X^1\Sigma^+_g\right)$, CO $\left(X^1\Sigma^+\right)$ and various electronic states of NO and  ICl diatomic molecules. 

Considering sDF molecular potential, Figure \ref{fig8} shows the variation of $E_{n,\ell}$ as a function of $\beta$, in the ground state. The restriction on choice of the parameter $\beta$ of sDF molecular potential can be observed. The energy for $\ell=1,2$ increases in the positive side but $\ell=0$ the energy increases in the negative side. On the other hand, in the first excited state, the energy increases in the negative side in the interval $0 <\beta < 15$ for $\ell=0,1,2$. 

Figure \ref{fig9} shows the variation of  $E_{n,\ell}$ as a function of reduced mass $\mu$. Small values of particle mass $\mu$ result into a sharp change in energy values for ground and first excited states for the orbital states. The energy becomes stable when $\mu>1$. This plot indicates how to choose or read the most reasonable values of $\mu$ which provides the most appropriate and not overlapping spectrum amongst orbital states. Figure \ref{fig10} is a plot of energy versus bond length $r_e$. The spectrum $E_{0,2} > E_{0,1} > E_{0,0}$ when $r_e > 0.2$ and $E_{1,2} > E_{1,1} > E_{1,0}$ when $r_e > 0.4$. Figure 10 set restrictions on the most suitable values of $r_e$. At $r_e<0.6-0.8fm$, the energy of different orbital states overlap and deteriorate sharply. 

The variation of the energy versus parameter $D$ is shown in Figure \ref{fig11}. For $n=0$ the energy increases and then decreases in the given range $0 <D < 25$ whereas for $n=1$, it is increasing in the same interval. Furthermore we generated the spectrum of several diatomic molecules using the sDF molecular potential for various states. The behavior of the plot energy against each potential parameter for various states provides us the most appropriate physical values for each parameter.  Figures 6 and 10 have different behaviors since they are a plots of energy against $r_e$ for two potentials. However, if one has set to choose large values for $r_e$  (say $r_e>0.6$) then the two curves will be similar.

Table \ref{Tab5} compares our results for H$_2$ and CO with those of the GPS method, Nikiforov-Uvarov method and AIM methods. Our currently found energy states are reasonably compared with the other findings. The vibrational energy is close to 7 digits with AIM \cite{BJ12} but found to agree with GPS \cite{RO1} up to 4 digits. However, the rotational-vibrational energy states are close close to 5 digits with AIM. This is due to the approximation made to the centrifugal restorsion term. Also, it should be noted that the model is a parameter dependent which may result into slight variation in energy spectrum if parameters are not converted/adjusted properly. Table \ref{Tab6} displays the energy spectrum for different species of $NO$ diatomic molecules. However, Table \ref{Tab7} present ones for H$_2$ $\left(X^1\Sigma^+_g\right)$, CO $\left(X^1\Sigma^+\right)$, various electronic states of  $ICl$ diatomic molecules

\section{Concluding Remarks}
In this research work, we applied proper quantization rule in a spectroscopic study of some diatomic molecules. This task is made possible by solving the Schr\"{o}dinger equation with two molecular models; namely, Tietz-Wei and shifted Deng-Fan potential models. This solution serves as the basis for the description of the quantum aspects of diatomic molecules. We obtained the energy spectra of different diatomic molecules. The validity and accuracy of the method is tested with previous techniques via numerical computation for H$_2$ and CO molecules. Our reasonable results show the efficiency and simplicity of the present calculations. The approximation to the centrifugal restorsion is valid for the lowest orbital quantum number $\ell$. As $\ell$ increases, the accuracy of the energy states reduces and vice-versa. The present research work represents a new procedure in dealing with the diatomic molecules. Our results are reasonable and credible in generating the spectrum as the other commonly known methods.

\section*{Acknowledgments}
We thank the kind referees for the positive enlightening comments and suggestions, which have greatly helped us in making improvements to this paper. In addition, BJF acknowledges eJDS (ICTP).


\begin{thebibliography}{99} 
\bibitem{NEQ1}B. J. Falaye, S. M. Ikhdair and M Hamzavi, Few Body Sys. {\bf56} (2015) 63
\bibitem{SM1}A. F. Nikiforov and V. B. Uvarov, Special Functions of Mathematical Physics, Berlin, Birkhauser, 1988.
\bibitem{SM2}H. \c{C}ift\c{c}i, R. L. Hall and N. Saad, J. Phys. A: Math Gen.  \textbf{36}(2003) 11807.\\
             H. \c{C}ift\c{c}i, R. L. Hall and N. Saad, Phys. Lett. A: \textbf{340}(2005) 388.
\bibitem{SM3}J. M. Fellows and R. A. Smith, J. Phys. A: Math. Theor. {\bf42} (2009) 333503.
\bibitem{SM4}S. H. Dong, Factorization Method in Quantum Mechanics, Springer, 2007.
\bibitem{SM5}S. M. Ikhdair, Adv. High Energy Phys. 2013 (2013) 491648.
\bibitem{P1} A. K. Roy, J. Phys. G: Nucl. Part. Phys. {\bf30} (2004) 269.\\
             A. K. Roy, Phys. Lett. A {\bf321} (2004) 231.\\
             A. K. Roy, J. Phys. B: At. Mol. Opt. Phys. 37 (2004) 4369.\\
             A. K. Roy, Int. J. Quant. Chem. {\bf104} (2005) 861.\\
             A. K. Roy, Int. J. Quant. Chem. {\bf108} (2008) 837.
\bibitem{SM7}Z. Q. Ma and B. W. Xu  Europhys. Lett. {\bf69} (2005) 685.\\
						 Z. Q. Ma and B. W. Xu  Int. J. Mod. Phys. E {\bf14} (2005) 599.
\bibitem{B1}S. H. Dong and A. Gonzalez-Cisneros, Ann. Phys. {\bf323} (2008) 1136.\\
						S. H. Dong, Int. J. Quant. Chem. {\bf109} (2009) 701.
\bibitem{B2}X. Y. Gu, S. H. Dong and Z. Q. Ma,  J. Phys. A: Math. Theor. {\bf42} (2009) 035303.
\bibitem{NEQ2}B. J. Falaye, S. M. Ikhdair and M Hamzavi, Phys. Scri. {\bf89} (2014) 115204.
\bibitem{SM10} F. A. Serrano, X. Y. Gu, S. H. Dong, J. Math. Phys. {\bf51} (2010)  082103.\\
							 F. A. Serrano and S. H. Dong, Int. J. Quantum Chem. 113 (2013) 2282.
\bibitem{SM11} S. H. Dong and M. Cruz-Irisson, J. Math. Chem. {\bf50} (2012) 881.
\bibitem{SM13} W. C. Qiang and S. H. Dong, EPL 89 (2010) 10003.
\bibitem{BE1}  C. Berkdemir, J. Math. Chem. 46 (2009) 492.
\bibitem{SM14} C. S. Jia et al, J. Chem. Phys. 137 (2012) 014101.
\bibitem{SM15} M. Hamzavi, S. M. Ikhdair and K. E. Thylwe, J. Math. Chem. 51 (2013) 227.
\bibitem{BJ5} C. N. Yang, {\it Monopoles in Quantum Field Theory}, Proceedings of the Monopole Meeting, Trieste, Italy, edited by N. S. Craigie, P. Goddard, and W. Nahm (World Scientific, Singapore, 1982), p. 237.
\bibitem{B6} S. H. Dong and X. Y. Gu, J. Phys: Conf. Series {\bf96} (2008) 012109.
\bibitem{B7} S. H. Dong, Commu. Theor. Phys. 55 (2011) 969.
\bibitem{BJ6} X. Y. Gu and S. H. Dong, J. Math. Chem. {\bf49} (2011) 2053.
\bibitem{BJ7} F. A. Serrano, M. Cruz-Irisson and S. H. Dong, Ann. Phys. (Berlin) {\bf523} (2011)  771.
\bibitem{BJ8} G. H. Sun and S. H Dong, Commun. Theor. Phys. {\bf58} (2012) 195.
\bibitem{BJ9} M. Hamzavi, A. A. Rajabi and H. Hassanabadi, Mol. Phys. {\bf110} (2012) 389.
\bibitem{BJ10} P. M. Morse, Phys. Rev. {\bf34} (1929) 57.
\bibitem{RO1} A. K. Roy. Int. J. Quant. Chem. (2013) DOI: 10.1002/qua.24575.
\bibitem{BJ12} K. J. Oyewumi, B. J. Falaye, C. A. Onate, O. J. Oluwadare and W. A. Yahya, Mol. Phys. {\bf112} (2014) 124.
\bibitem{BJ13} P. Q. Wang, L H Zhang, C. S. Jia and J. Y. Liu, J. Mol. Spect. 274 (2012) 5
\bibitem{BJ14} F. J. Gordillo-Vizquez, J. A. Kunc, J.  Appl.  Phys. {\bf84} (1998) 4693.
\bibitem{AKR1} A. K. Roy, J. Math. Chem. {\bf52} (2014) 1405.
\bibitem{BJ15} S. M. Ikhdair and B. J. Falaye, Chem. Phys. {\bf421} (2013) 84.
\bibitem{BJ16} The NIST reference on constants, units and uncertainty, physics.nist.gov/cuu/Constants/index.html.
\end{thebibliography}
\end{document}